\documentclass{article}

\begin{document}

\newtheorem{defn}{Definition}
\def\thedefn{\thesection.\arabic{defn}}
             
\newtheorem{teor}[defn]{Theorem}
\newtheorem{ejem}[defn]{Example}
\newtheorem{lema}[defn]{Lemma}
\newtheorem{rema}[defn]{Remark}
\newtheorem{coro}[defn]{Corollary}
\newtheorem{prop}[defn]{Proposition}

\makeatother
\font\ddpp=msbm10  
\def\R{\hbox{\ddpp R}}     
\def\C{\hbox{\ddpp C}}     
\def\L{\hbox{\ddpp L}}    
\def\S{\hbox{\ddpp S}}
\def\Z{\hbox{\ddpp Z}}
\def\Q{\hbox{\ddpp Q}}     
\def\N{\hbox{\ddpp N}}

\newcommand{\dps}{d_{\psi}}
\newcommand{\df}{d_{\phi}}

\newcommand{\M}{{\cal M}}
\newcommand{\Mo}{{\cal M}_0}
\newcommand{\be}{\begin{equation}}
\newcommand{\ee}{\end{equation}}
\newcommand{\la}{\Lambda}
\newcommand{\dem}{\noindent {\rm Proof: }}
\newcommand{\inte}{\int_{0}^{1}}
\newcommand{\gam}{\gamma}
\newcommand{\eps}{\epsilon}
\newcommand{\<}{\langle}
\renewcommand{\>}{\rangle}
\newcommand{\Om}{\Omega^1}
\renewcommand{\(}{\left(}
\renewcommand{\)}{\right)}
\renewcommand{\[}{\left[}
\renewcommand{\]}{\right]}
\newcommand{\om}{\omega}
\newcommand{\me}{\frac{1}{2}}
\newcommand{\Mt}{\widetilde{\M}}
\newcommand{\cat}{{\mathop{\rm cat}\nolimits}}
\newcommand{\Tau}{{\cal T}}

\newcommand{\cvd}{{\rule{0.5em}{0.5em}}\smallskip}

\hyphenation{Lo-rent-zian}

\title{{\bf Smoothness of time functions \\ and the metric splitting of \\ globally hyperbolic  spacetimes}}
\author{Antonio N. Bernal  and Miguel S\'anchez\thanks{
The second-named author has been partially supported by a  MCyT-FEDER Grant, MTM2004-04934-C04-01.
}}
\maketitle

\begin{center}
{\small Dpto. de Geometr\'{\i}a y Topolog\'{\i}a, Facultad de Ciencias, Fuentenueva s/n, E--18071 Granada, Spain}
\end{center}

\begin{flushright}
To Professor P.E. Ehrlich,  wishing him \\
a continued recovery and good health
\end{flushright}

\begin{center}
Abstract
\end{center}

\noindent The folk questions in Lorentzian Geometry which  concerns the smoothness of time functions and slicings by Cauchy hypersurfaces, are  solved by giving simple proofs of: (a) any globally hyperbolic spacetime $(M,g)$ admits a smooth time function $\Tau$ whose levels are spacelike Cauchy hyperfurfaces and, thus, also a smooth 
global splitting $M= \R \times {\cal S}$, $g= - \beta(\Tau,x) d\Tau^2 + \bar g_\Tau $, (b) if a spacetime $M$ admits a (continuous) time function $t$  then it admits a smooth (time) function $\Tau$ with timelike gradient $\nabla \Tau$ on all $M$. 


\smallskip

\section{Introduction}

The present article deals with some folk questions on differentiability of time functions and Cauchy hypersurfaces, as a natural continuation of our previous paper \cite{BS}. The following questions have been widely controversial since the very beginning of Causality Theory (see \cite[Section 1]{BS} for a discussion and references\footnote{See also the authors' contribution to {\em Proc. II Int. Meeting on Lorentzian Geometry}, Murcia, Spain, 2003, Publ. RSME vol. 8 (2004) 3--14, gr-qc/0404084, where the present results have been announced.}): (i) must  any globally hyperbolic spacetime contain  a {\em smooth spacelike} Cauchy hypersurface? \cite[p. 1155]{SW} (ii) can classical Geroch's topological splitting of globally hyperbolic spacetimes \cite{Ge} be strengthened in a {\em smooth orthogonal splitting}?, and (iii) does any stably causal spacetime admit a {\em smooth}  function with {\em timelike} gradient on all $M$? \cite[p. 64]{BEE}. The first question was answered affirmatively in \cite{BS}, and our aim  is to answer the other two.

Concretely, for question (ii) we prove:

\begin{teor} \label{t0} 
Let $(M,g)$ be a globally hyperbolic spacetime. Then, it is isommetric to the smooth product manifold
$$
\R \times {\cal S}, \quad \langle \cdot , \cdot \rangle = - \beta\,d\Tau^2 + \bar g 
$$
where ${\cal S}$ is a smooth spacelike Cauchy hypersurface, $\Tau: \R\times {\cal S} \rightarrow \R$ is the natural projection, $\beta:\R \times {\cal S} \rightarrow (0,\infty)$ a smooth function, and $\bar g$ a 2-covariant symmetric tensor field on $\R \times {\cal S}$, satisfying:

\begin{enumerate}
\item $\nabla \Tau$ is timelike and past-pointing on all $M$ (in particular, $\Tau$ is a time function).

\item Each hypersurface ${\cal S}_\Tau$ at constant $\Tau$ is a Cauchy hypersurface, and 
the restriction $\bar g_{\Tau}$ of $\bar g$ to such a ${\cal S}_\Tau$ is a Riemannian metric (i.e. ${\cal S}_\Tau$ is spacelike).

\item The radical of $\bar g$ at each $w\in \R \times {\cal S}$ is Span$\nabla \Tau$ (=Span $\partial_\Tau$) at $w$.

\end{enumerate}
\end{teor}
For question (iii), recall first that  a stably causal spacetime $M$ is a causal spacetime which remains causal when its metric is varied in some  neighborhood for the $C^0$ topology of metrics \cite[p. 242]{BEE}.  It is well-known from the causal ladder of spacetimes  that any globally hyperbolic spacetime is stably causal, but the converse does not hold \cite[p. 73]{BEE}; even more,  stably causal spacetimes may fail to be causally continuous (\cite[p. 71]{BEE}, see also \cite{Sa} for detailed proofs and discussions on causally continuous and stably causal spacetimes). Hawking  \cite{Ha} proved that any stably causal spacetime admits a time function, i.e., a {\em continuous} function $t: M \rightarrow \R$ which is strictly increasing on any future--directed causal curve. In fact, causally continuous spacetimes are characterized as spacetimes such that the past and future volume functions are time functions (for one, and then for any, admissible Borel measure). In the case of stably causal spacetimes, a time function is obtained as an appropiate ``average'' of volume functions for causal metrics obtained by widening the cones of the original one. Nevertheless,  even though  the continuity of such an average function is proved, its smoothability remained as a ``folk problem''  (see also \cite[Proposition 6.4.9]{HE} , \cite[Section 3.2]{BEE}, \cite[Sect. 4]{Sa}). Conversely, Hawking also proved that a spacetime is stably causal if it admits 
a smooth function with everywhere timelike gradient. 

We will call such a function a {\em temporal function}, i.e., a smooth function $\Tau$ on a spacetime $M$ with (past-directed) timelike gradient $\nabla \Tau$ on all $M$. Notice that, obviously, any temporal function  is a time function, but even a smooth time  function may be non-temporal (it may have a lightlike  gradient in some points). Our technique also proves:
\begin{teor} \label{t0b}
Any spacetime $M$ which admits a time function also admits a temporal function. 
\end{teor}
This result, combined with Hawking's ones, ensures, on one hand, that any stably causal spacetime admits a temporal function and, on the other, that any spacetime which admits a time function is stably causal.

Notice that, in fact, Theorem \ref{t0} ensures the existence of a {\em Cauchy temporal function} $\Tau$ in any globally hyperbolic spacetime, i.e.,  a temporal function with Cauchy hypersurfaces as levels. So, the proof of Theorem \ref{t0b} can be carried out easily by simplifying the  reasonings for Theorem \ref{t0} (the property of being Cauchy is not taken into account now). For the proof of this last theorem, the reader is assumed to be familiarized with the technique  in \cite{BS}. 

Our approach is very different to previous ones on this topic. Essentially, the idea goes as follows. Let $t$ be a continuous Cauchy time function as in Geroch's theorem.
As shown in \cite{BS}, if $t_-<t$ then there exists a smooth Cauchy hypersurface ${\cal S}$ contained in    $t^{-1}(t_-, t)$; this hypersurface is obtained as the regular value of certain function with either timelike or zero gradient on $t^{-1}(t_-, t]$. As $t_-$ approaches $t$, ${\cal S}$ can be seen as a smoothing of $S_{t}$; nevertheless ${\cal S}$ always lies in $I^-(S_t)$. In Section \ref{s2} we show how the required splitting of the spacetime would be obtained if we could ensure the existence of  a {\em temporal step function} $\tau$ around each $S_t$. Essentially, such a $\tau$ is a function with timelike gradient on a neighborhood of $S_t$ (and zero gradient outside) with level Cauchy hypersurfaces which cover a rectangular neighbourhood of $S_t$ (Definition \ref{d}). Then, Section \ref{s3} is devoted to prove the existence of such a temporal step function around any $S_t$. To this aim, we will show first how $S_t$ (but perhaps no other close Cauchy hypersurface obtained varying $t$) can be covered by Cauchy level hypersurfaces of a certain function $\hat \tau$, Proposition \ref{p1}. Then, the temporal step function $\tau$ will be obtained as the sum of a series of previously constructed functions $\sum_j \hat \tau^{[j]}$, Theorem \ref{t1}, and special care will be necessary to ensure its smoothness.  

Finally, in Remark \ref{r} we sketch Theorem \ref{t0b}. Essentially, only some variations on previous arguments are needed, and we also  sketch a proof by taking into account that each level hypersurface $S$ of the continuous time function $t$ is not only achronal, but also a Cauchy hypersurface in its Cauchy development $D(S)$.
 
\section{Temporal step functions }\label{s2}

In what follows, $M\equiv (M,g)$ will denote a $n$-dimensional globally hyperbolic spacetime, and $t$  a continuous Cauchy time function given by Geroch's theorem, as in \cite[Proposition 4]{BS}. Then,  each $S_t = t^{-1}(t)$ is the corresponding topological Cauchy hypersurface, and the associated topological splitting is $M=\R\times S$, where $S$ is any of the $S_t$'s (see \cite[Proposition 5]{BS}). 

$\N= \{1,2,\dots \}$ will denote the natural numbers ($\Z , \R$, resp.,   integers,  real numbers). 
In principle, ``smooth'' means $C^r$-differentiable, where $r\in \N \cup \{ \infty \}$ is the maximum degree of differentiability of the spacetime. Nevertheless, we will assume $r= \infty$, and the steps would remain equal if $r<\infty$, except for some obvious simplifications in the proof of Theorem \ref{t1}.  $\overline{W}$  will denote the topological closure of the subset $W\subset M$.

\begin{defn} \label{dts}  
\label{d} {\rm Given the Cauchy hypersurface $S \equiv S_{t}$, fix $t_-, t_+, t_{a}, t_b \in \R$, $t_- < t_{a} < t < t_b < t_+$, and put $S_- = S_{t_-}, S_+= S_{t_+}$, $I_t=(t_{a},t_b)$. We will say that 
$$\tau : M\rightarrow \R$$
is a {\em temporal step function around $t$}, compatible with the outer extremes $t_-, t_+$ and the inner extremes $t_{a}, t_b$,  if it satisfies:
 
\begin{enumerate}

\item $\nabla \tau$ is timelike and past-pointing where it does not vanish, that is, in the interior of its support
$V :={\rm Int(Supp}(\nabla \tau )$).

\item $-1 \leq \tau \leq 1$.

\item $\tau(J^+(S_{+})) \equiv 1$, $\tau(J^-(S_{-})) \equiv -1$. In particular, the support of $\nabla \tau$ satisfies: ${\rm Supp}(\nabla \tau) \subset J^+(S_-) \cap J^-(S_+) (=t^{-1}[t_-,t_+]).$

\item $S_{t'} \subset V $, for all $t' \in I_{t}$; that is, the rectangular neighborhood of $S$, $t^{-1}(I_t) \equiv I_t \times S$, is included in $V$ (or $J^+(S_{t_{a}}) \cap J^-(S_{t_b})  \subset {\rm Supp}(\nabla \tau)$).

\end{enumerate}
}
\end{defn}
Recall that, from the first property,  the inverse image of any regular value of $\tau$ is a smooth closed\footnote{Here, {\em closed} hypersurface means {\em closed as a subset of $M$} (but not necessarily compact); hypersurfaces are always assumed embedded without boundary.}  spacelike hypersurface. Even more, from the third property such hypersurfaces are Cauchy  hypersurfaces  (use \cite[Corollary 11]{BS}) and, from the fourth, they cover not only $S_t$ but also close Cauchy hypersurfaces $S_{t'}$. 

\begin{prop} \label{p0} 
Assume that the globally hyperbolic spacetime $M$ admits a temporal step function around $t$, for any $t \in \R$, compatible with outer extremes $t_+= t+2, t_-= t-2$. Then, there exists a smooth function $\Tau: M\rightarrow \R$ which satisfies:

(A) $\nabla \Tau$ is timelike and past-pointing on all $M$.

(B) For each inextendible timelike curve $\gamma: \R \rightarrow M$, parameterized with $t$, 
one has:
lim$_{t\rightarrow \pm \infty}(\Tau(\gamma(t)) = \pm \infty$.

Thus, each hypersurface at constant $\Tau $,  ${\cal S}_\Tau = \Tau^{-1}(\Tau)$, is a smooth spacelike Cauchy hypersurface, and all the conclusions of Theorem \ref{t0} holds.
\end{prop}

\noindent {\em Proof.} Consider, for each $t\in \R$, the function $\tau \equiv \tau_t$, the open subset $V_t$ and the interval $I_t$ in Definition \ref{d},  and, thus, the open covering of $M$
$${\cal V} = \{V_t: t\in \R\} $$
with associated open covering ${\cal I}= \{I_t , t\in \R\}$ of $\R$. As the length of each $I_t$ is $< t_+ - t_-=4$, a locally finite subrecovering ${\cal I}'$ of ${\cal I}$ exists (we can also assume that no interval in this subrecovering is included in another interval of it) and, as a consequence, a locally finite subrecovering ${\cal V'}$
of ${\cal V}$:

$${\cal V'} = \{V_{t_k}: k \in \Z \}. $$

\noindent Without loss of generality, we can assume $t_k < t_{k+1}$ and then, necessarily, 
\be \label{e1}
\lim_{k \rightarrow \pm \infty} t_k = \pm \infty .
\ee
The notation will be simplified $V_k \equiv V_{t_k},   \tau_k \equiv \tau_{t_k}$. 

Define now:
\be \label{e1.5}
 \Tau = \tau_0 + \sum_{k=1}^{\infty} (\tau_{-k} + \tau_k ) .
\ee
Notice that $\tau_{-k} + \tau_k \equiv 0 $ on 
$J^+(S_{t_{-k}+2}) \cap J^-(S_{t_{k}-2})$, 
for all $k$ (this applies when $k>k_0$, where $k_0$ is the first  $k>0$ such that $t_{-k}+2  < t_{k}-2$). This (plus the limit (\ref{e1})) ensures that $\Tau$ is well defined and smooth.

Property (A) is then straightforward from the definition of $\Tau$, the convexity of the (past) time cones and the fact that  ${\cal V'}$ covers all $M$.

For (B), consider the limit to $+\infty$ (to $-\infty$ is analogous). It is enough to check that, for any $k \in \N$ there exists $t^k \in \R$ such that  $\Tau(\gamma(t^k)) > k$ (and, thus, from (A), this inequality holds for all $t>t^k$). But taking $t^k= t_k+2$ ($\geq {\rm Sup}(t(V_k))$), one has $\Tau(\gamma(t^k)) > 2k$ obviously from (\ref{e1.5}).

To check that the (necessarily smooth and spacelike) hypersurface $S_\Tau$ is Cauchy, notice that no timelike curve (in fact, no causal one) can cross more than once ${\cal S}_\Tau$ because of property (A). Thus, any inextendible timelike curve $\gamma$ can be  $\Tau$-reparameterized in some interval $(\Tau_-, \Tau_+)$ and, because of (B), necessarily $\Tau_\pm = \pm \infty$. Therefore, $\gamma$ must cross each $S_\Tau$.  

Now, the assertions in Theorem \ref{t0} are straightforward consequences of previous properties. Briefly, let
${\cal S}= \Tau^{-1}(0)$ and define the map 
$$ \Phi: M \rightarrow \R \times {\cal S} , \quad \quad p \rightarrow (\Tau(p), \Pi(p)),$$ where
 $\Pi(p)$ is the unique point of  ${\cal S}$ crossed by the inextendible curve of  
$\nabla \Tau$ through $p$. The vector field  $\partial/\partial\Tau$ obtained at each point  $p$ as the derivative of the curve $s\rightarrow \Phi^{-1}(\Tau(p)+s, \Pi(p))$ is clearly colinear to $\nabla \Tau$ at each point. Even more, as $g(\partial/\partial\Tau,\nabla \Tau) \equiv 1$, then  $\partial/\partial\Tau = -\nabla \Tau/|\nabla \Tau|^2$. Thus, the metric $\Phi^*g$ induced on $\R \times {\cal S}$ satisfies all the required properties with $\beta (\Phi(p))= |\nabla \Tau|^{-2}(p)$, for all $p\in M$.  \cvd

\begin{rema} {\rm 
(i) The  restriction on the outer extremes can be obviously weakened by assuming that $t_+-t_-$ is bounded. Thus, Proposition \ref{p0} reduces Theorem \ref{t0} to prove the existence of a temporal step function around any $S_t$ with bounded outer extremes. Theorem \ref{t1} will prove this result.

(ii) In fact, Theorem \ref{t1} proves more: the outer and inner extremes can be chosen arbitrarily.  Thus, one can assume always for temporal step functions $t_b= t+1, t_{a} = t-1, t_+ = t+2, t_- = t-2$. In this case, the proof of  Proposition \ref{p0} can be simplified because one can take directly the subrecovering ${\cal V}'$ with $t_k = k $ for all $k\in \Z$. 
}
\end{rema}
\section{Construction of a temporal step function } \label{s3}

\begin{prop} \label{p1}
For each $S \equiv S_{t}$  there exists a function $\hat \tau\equiv \hat \tau_t$ which satisfies the three first properties  in Definition \ref{d} and, additionally:

$\hat 4$. $S\subset V$.

\end{prop}
For its proof, we will need first the following two lemmas, which are straightforward from  \cite{BS}. Thus, we only sketch the steps for their proofs.

\begin{lema} \label{p1l1}
Let $S\equiv S_t$ be a Cauchy hypersurface. Then there exists an open subset $U$ with
$$J^-(S) \subset U \subset I^-(S_{t+1}) ,$$  
and a function 
$h^+: M \rightarrow \R, h^+\geq 0, $ with support included in 
$I^+(S_{t-1})$ which satisfies: 

(i) If $p\in U$ with $h^+(p) >0$ then $\nabla h^+(p)$ is timelike and past-pointing.

(ii) $h^+ > 1/2 $ (and, thus, its gradient is timelike past pointing) on $J^+(S)\cap U$.
 
\end{lema}

\noindent {\em Proof.} 
Recall that the function $h$ constructed in 
 \cite[Proposition 14]{BS} (putting $S=S_2$, $S_{t-1}=S_1$) yields directly  $h^+$. In fact,  function $h$ in that reference also satisfies {\em (ii)} in an open neighborhood $U'$ of $S$ (this is obvious because that function $h$ is constructed from the sum of certain functions $h_p$ in \cite[Lemma 5]{BS} which satisfy {\em (ii)} on appropiate open subsets which cover $S$), which can be chosen included in $I^-(S_{t+1})$, 
 and, thus, take $U= U' \cup I^-(S)$. \cvd

\begin{lema} \label{p1l2}
Let $S$ be a Cauchy hipersuperface and $U (\subset I^-(S_{t+1}))$ an open neighborhood of $J^-(S)$.  Then, there exists a function
$h^-: M \rightarrow \R, h^-\leq 0, $ with support included in $U $ satisfying: 

(i) If $\nabla h^-(p)\neq 0$ at $p \, (\in U)$ then $\nabla h^-(p)$ is timelike past-pointing.

(ii) $h^- \equiv \, -1$ on $J^-(S)$.

\end{lema}

\noindent {\em Proof.} The proof would follow as the construction of  $h$ in \cite[Proposition 14]{BS} with the following modifications: 
(a) reverse the time-orientation, and consider  \cite[Proposition 14]{BS} with $S=S_2$, $S_{t+1}=S_1$, (b) take all the convex open subsets ${\cal C}_p$ included in $U$, (c) construct $h$ by exactly the same method, but changing the sign of all the Lorentz distances (i.e., time--separations are taken negative), and (d) once $h\leq 0$   is constructed in this way (notice that $h(S)<-1/2$), define $h^-$ on $J^+(S)$ (with the original time-orientation in what follows)
as:
$$h^-=\varphi \circ h ,$$ where  $\varphi: \R \rightarrow \R$ is any function which satisfies $$\varphi((-\infty,-1/2]) \equiv \, -1 , \quad \varphi'([-1/2,0]) >0, \quad \varphi(0)=0$$
(of course, $h^-$ is defined on $J^-(S)$ as equal to $-1$).
\cvd

\smallskip

\smallskip

\noindent {\em Proof of Proposition \ref{p1}}.  
Fixed $S_t$, take $U$, $h^+$ and $h^-$ as in the two previous lemmas. Notice that 
$h^+ - h^- >0$ on all $U$. Then, define: 
$$
\hat \tau_t = 2 \; \frac{h^+}{h^+ - h^-} -1
$$
on $U$, and constantly equal to 1 on $M\backslash U$.
As
$$ \nabla \hat \tau_t = 2 \; \frac{h^+ \nabla h^- - h^- \nabla h^+}{(h^+ - h^-)^2} 
$$
is either timelike or 0 everywhere, all the required properties are trivially satisfied. \cvd

\smallskip

\smallskip

\noindent We can even strengthen technically the conclusion of Proposition \ref{p1} for posterior referencing:
\begin{coro} \label{c1}
Let $t_-< t_{a} < t < t_b < t_+$ and a compact subset $K \subset t^{-1}([t_{a}, t_b]) $
be. Then, there exists a function $\hat \tau$ which satisfies the four properties of Proposition \ref{p1} and, additionally:
$K \subset V.$
\end{coro}

\noindent {\em Proof.} For each $S_t$ with $t\in [t_a, t_b]$, take the corresponding function $\hat \tau_t$ from Proposition \ref{p1}.  $K$ is then covered by the corresponding open subsets $V_t$ and, from compactness,  a finite set of $t$'s, say, $t_1, \dots , t_m$ suffices. Then take $\hat \tau= m^{-1} \sum_i \hat \tau_{t_i}$. \cvd

\noindent Theorem \ref{t0} will be the obvious consequence of Proposition \ref{p0} and Theorem \ref{t1} below. For the proof of this one, we will sum an appropiate series of functions as the ones in Corollary \ref{c1}, and we will have to be careful with the smoothness of the sum. But, first, the following trivial lemma will ensure that  the  infinite sum will not be an obstacle for the timelike character of the gradient. 

\begin{lema} \label{lz} Let $\{ v_i \}$ be a sequence of timelike vectors in the same cone of a vector space. If the sum $v= \sum_{i=1}^\infty  v_i$ is well defined then the vector $v$ is timelike.
\end{lema}

\noindent {\em Proof.} As the causal cones are closed, $\sum_{i=2}^\infty  v_i$ is causal and, as the sum of a causal plus a timelike vector in the same cone is timelike, 
$v = v_1 +  \sum_{i=2}^\infty v_i$ is timelike. \cvd

\begin{teor} \label{t1}
For each $S \equiv S_{t}$ and $t_-<t_{a}<t<t_b<t_+$ there exists a temporal step function $\tau$ around $S$ with outer extremes $t_- , t_+$ and inner extremes $t_{a}, t_b$, $I_t=(t_{a},t_b)$.

\end{teor}

\noindent {\em Proof.} Choose a sequence $\{G_j: j \in \N \}$ of open subsets such that:

\be \label{eg}
\overline{G_j}\,\,\,\mbox{is compact},\,\,\,\overline{G_j}\subset G_{j+1}
 \,\,\,\mbox{}\,\,\,M=\cup_{j=1}^\infty G_j , 
\ee
and the associated sequence of inner compact subsets
$$
K_j=\overline{G_j}\cap J^+(S_{t_{a}})\cap J^-(S_{t_b}).$$ 
For each $K_j$, consider the function $\hat \tau^{[j]}$ given by Corollary \ref{c1} with $K=K_j$, and put $V_j:= $Int(Supp $\nabla \hat \tau^{[j]})$, $K_j \subset V_j$. Notice that the series
\be \label{es}
\tilde \tau:= \sum_{j=1}^\infty \frac{1}{2^j } \hat \tau^{[j]},
\ee
converges at each $q\in M$ and, thus, defines a continuous function $\tilde \tau: M \rightarrow \R$. If $\tilde \tau$ were smooth and its partial derivatives  (in coordinate charts) conmuted with the infinite $\sum$, then  $\tilde \tau$ would be the required temporal step function, obviously (use Lemma \ref{lz}). As  these hypotheses have not been ensured, expression (\ref{es}) will be modified as follows.

Fix a locally finite atlas  ${\cal A} = \{W_i: i\in \N \} $  such that each chart $W \equiv (W, x_1, \dots , x_n)\in {\cal A}$ has a relatively compact domain and it is also the restriction of a bigger chart on $M$ whose domain includes $\overline{W}$. Then, each compact subset $\overline{ G_j}$ is intersected by a finite number of neighborhoods $W_{i_1}, \dots , W_{i_{k_j}}$. As  $D:= \overline{(W_{i_1} \cup \dots \cup W_{i_{k_j}}}$) is compact, there exists  $A_j >1$  such that
$|\hat \tau^{[j]}| < A_j$ on $D$ and, for each $s<j$:
$$ 
\left| 
 \frac{\partial^s \hat \tau^{[j]}}{\partial x_{l_1}\partial x_{l_2}...\partial x_{l_s}}(q) \right| <A_j , \quad \quad \forall q\in D , \forall l_1, \dots , l_s \in \{1,\dots , n\}.
$$
Now, the series
\be \label{es2}
\tau^*:= \sum_{j=1}^\infty \frac{1}{2^j A_j} \hat \tau^{[j]},
\ee
is smooth on all $M$. In fact, to check differentiability $C^s$ at $p\in M$, choose $j_0 \in \N$ and $W \in {\cal A}$ with $p \in G_{j_0} \cap W$. Recall that, for any $j > $ Max$\{j_0, s\}$, the summand $\frac{1}{2^j A_j} \hat \tau^{[j]}$ and all its partial derivatives in the local coordinates of $W$ until order $s$, are bounded in absolute value by $1/2^j$ on $G_{j_0} \cap W$. Thus, the series (\ref{es2}) and the partial derivatives converge uniformly on a neighborhood of $p$, and the derivatives conmute with $\sum$ on $M$. 

Therefore, $\tau^*$ satisfies trivially all the properties of a temporal step function in Definition \ref{d} except, at most, the normalizations to 1 and -1 in the second and third ones. Instead, $\tau^*$ satisfies $\tau^*(J^-(S_-)) \equiv c_-<0, \tau^*(J^+(S_+)) \equiv c_+>0. $
The required function is then $\tau = \psi \circ \tau^*$, where the smooth function $\psi : \R \rightarrow \R$ satisfies $\psi'>0, \psi(c_-) = -1, \psi(c_+)=1$. \cvd

\begin{rema} \label{r} {\rm 
As said in the Introduction, the proof of Theorem \ref{t0b} can be carried out directly by simplifying previous reasonings. Concretely, property (B) of Proposition \ref{p0} is not needed now, and property (A) can be achieved from temporal step functions as in Definition \ref{dts}, where each $S_t$ is a level hypersurface of the time function.
Alternatively, let $t$ be a time function, choose $p \in M$ and let $S=t^{-1}(t(p))$ be the level hypersurface of $t$ through $p$. Then, $S$ is closed, achronal and separates $M$, i.e., $M\backslash S$ 
 is the disjoint union of the open subsets, $M_+ := t^{-1}(t(p), \infty) (\supseteq I^+(S))$ and $M_-:= t^{-1}(-\infty, t(p)) (\supseteq I^-(S))$. Even more, $S$ is a Cauchy hypersurface of its Cauchy development $D(S)$. Now, recall: 

(i) Any temporal step function $\tau$ on $D(S)$ can be extended to all $M$ by putting $\tau(M_+\backslash D^+(S)) \equiv 1$, 
$\tau (M_-\backslash D^-(S)) \equiv -1$. Thus, $\nabla \tau$ is: (a) either timelike or null everywhere, and (b) timelike on a neighborhood of $p$. 

(ii) Given any compact subset $G \subset M$, a similar function $\hat \tau$, which satisfies not only (a) but also (b) for all $p \in G$,  can be obtained as a finite sum of functions constructed in (i) (in analogy to Corollary \ref{c1}).

(iii) Choosing a sequence of compact subsets $G_j$ as in formula (\ref{eg}), taking the   corresponding function $\hat \tau^{[j]}$ obtained in (ii), and summing a series in a similar way than in (\ref{es2}), the required $\tau$ is obtained.  
}\end{rema}

\end{document}